\documentclass[12pt,a4paper]{article}
\usepackage[cp866]{inputenc}
\usepackage[T2A]{fontenc}
\usepackage{amsfonts,amsmath,amssymb,amscd,euscript}
\usepackage[russian,english]{babel}
\normalbaselines
\newcommand{\be}{\begin{equation}\label}
\newcommand{\ee}{\end{equation}}
\newcommand{\prt}{\partial}
\newcommand{\p}{\prime}
\newcommand{\bib}{\bibitem}
         
\newcommand{\wh}{\widehat}          
              
\begin{document}

\begin{center} 

{\bf  TWISTOR ALGEBRAIC DYNAMICS IN COMPLEX SPACE-TIME \\ 
AND PHYSICAL MEANING OF HIDDEN DIMENSIONS}

\bigskip

{\bf Vladimir V. Kassandrov}

\end{center}

\bigskip

\noindent
{\footnotesize {\bf Abstract.} We review the algebraic field theory 
based completely on a nonlinear generalization of the CR complex analiticity 
conditions to the noncommutative algebra of biquaternions. Any biquaternionic 
field possesses natural twistor structure and, in the Minkowski space, 
gives rise to a shear-free null congruence of rays and to a set of 
gauge fields associated with it. In the paper we develop the algebrodynamical 
scheme in the complex extension of Minkowski space-time -- in 
the full vector space of biquaternion algebra. This primodial space dynamically 
reduces to the 6D ``observable'' space-time of the complex null cone which 
in turn decomposes into the 4D physical space-time and the 2D internal 
``spin space''. A set of identical point charges (``duplicons'') -- focal points 
of the congruence -- arises in the procedure. Temporal dynamics of 
individual duplicons is strongly correlated via fundamental twistor field of 
the congruence. We briefly discuss some new notions inevitably arising in the 
considered algebrodynamical scheme, namely those of ``complex time'' and of 
``evolutionary curve'', as well as their hypothetical links to the  
quantum uncertainty phenomenon. 

} 

\section{Introduction}\label{Introd}

In the article we continue to develop the {\it algebrodynamical} field theory  
which has been presented, in particular, in~\cite{PIRT,Pavlov}. We recall that 
in the paradigm of algebrodynamics one considers, in the spirit of radical 
Pithagorean philosophy, all fundamental physical laws and phenomena as a 
manifestation of some basic abstract ``World'' structure -- {\it the Code of 
the Universe} -- which might be an exceptional group (e.g., the {\it Monster 
group}), algebra or geometry. The whole structure and evolution of the Universe, 
the cathegories of time, particle, field, motion and interaction should be all 
predetermined and encoded in internal properties of the World structure. 
The known physical laws are nothing but separate {\it fragments} of this 
basic structure, and one could try to {\it forget} them all and to read 
{\it unprejudicely} the ``Book of Nature'' which could bring us by this 
to a grand and magestic picture of physical World {\it quite different from 
that accepted in the modern theoretical physics}. 

In the main version of algebrodynamics developed by the author (see, e.g.,
~\cite{AD,GR,Acta,Sing,GRG} and references therein) the exceptional algebra of 
{\it quaternions} (precisely, of their complexification -- the algebra of 
{\it biquaternions} $\mathbb{B}$) is considered as the {\it World algebra}.  
Basic physical fields are then $\mathbb{B}$-valued functions of 
$\mathbb{B}$-variable, and field equations are nothing but the 
{\it conditions of $\mathbb{B}$-differentiability}, 
i.e. the {\it generalized Cauchy-Riemann equations} (GCRE). These have been 
introduced by the author in 1980 (see in detail~\cite{AD,GR,Est,Joseph,Protvino}) 
and have the following invariant form: 
\be{GCRE}
dF = \Phi * dZ * \Psi , 
\ee
where $F: \mathbb{B} \mapsto \mathbb{B}$ is a $\mathbb{B}$-differentiable 
function $F(Z)$ of $\mathbb{B}$-variable $Z\in \mathbb{B}$;~  
$\Phi,\Psi: \mathbb{B} \mapsto \mathbb{B}$ are two auxiliary functions 
(left and right ``semiderivatives'') associated with $F$, and $(*)$ 
denotes the operation of multiplication in $\mathbb{B}$ isomorphic to the full 
$2\times 2$ complex matrix algebra. 

As a direct consequence of (\ref{GCRE}), any matrix component of the 
$\mathbb{B}$-field $f=\{F_{AB}\},~A,B=1,2$  obeys the determinant-type equation  
\be{equat}
\det \Vert \frac{\prt f}{\prt Z_{AB}}\Vert = 0 
\ee
which, for the case of noncommutative $\mathbb{B}$-algebra, is a {\it nonlinear} 
analogue of the Laplace {\it harmonicity} equation in complex analysis.  

Let us restrict now the coordinates $Z$ onto the subspace $\bf M$ of the full 
$4\mathbb{C}$ vector space of $\mathbb{B}$-algebra represented by Hermitian 
matrices $X=X^+$. This has the metrical structure identical to that of 
{\it Minkowski} space. Below we'll use the following representation for 
the coordinates on $\bf M$:
\be{restr}
Z \mapsto X = 
\left(
\begin{array}{cc}
u & w \\
p & v 
\end{array}\right)
\ee
where $u=ct+z,~v=ct-z$ are real, $w=x-iy$ and $p=x+iy$ -- complex conjugated 
and $x,y,z,t$ -- Cartezian and time coordinates respectively. Now the GCRE 
read as follows:
\be{GSE}
dF = \Phi * dX * \Psi , 
\ee
and the fundamental equation (\ref{equat}) takes the form 
\be{CEE}
\frac{\prt f}{\prt u}~\frac{\prt f}{\prt v}-\frac{\prt f}{\prt w}~\frac{\prt f}
{\prt p}~\equiv~\frac{1}{c^2}\left(\frac{\prt f}{\prt t}\right)^2-\left(\frac{\prt f}{\prt x}
\right)^2-\left(\frac{\prt f}{\prt y}\right)^2-\left(\frac{\prt f}{\prt z}
\right)^2 = 0
\ee
in which one easily recognizes the {\it complex eikonal equation} (CEE)~\cite{AD}. 

On the Minkowski coordinate background the whole theory turns to be Lorentz 
invariant, and one is able to construct an ``algebraic'' field theory in which  
the nonlinear complex eikonal is a basic {\it self-interacting} physical field.
 
Principles and results of (nonlinear, non-Lagrangian, overdetermined) algebraic 
$\mathbb{B}$-field theory have been presented in a series of papers (see, e.g.,
~\cite{PIRT,Pavlov,GR,Joseph,Protvino,Kasan,Vestnik,Eik,GRG} and other papers). 
The GCRE and the related 
CEE were found to possess 2-spinor, twistor and restricted gauge structures. 
Moreover, {\it the integrability conditions for (\ref{GSE}) result in identical 
satisfaction of the (complexified) homogeneous Maxwell and Yang-Mills equations}. 

In our paper~\cite{Eik} the Lorentz invariant form of {\it general solution} to 
the CEE has been found on the base of its (ambi)twistor structure. Precisely, 
any (almost everywhere analytical) solution of CEE can be obtained from some 
generating function 
\be{genf}
\Pi(G,\tau^1,\tau^2) \equiv \Pi(G,wG+u,vG+p)
\ee
of three complex variables -- components of the projective (null) twistor related 
to the points of space-time $X\in \bf M$ via the Penrose {\it incidence condition}
~\cite{PR}
\be{inc}
\tau = X\xi~~(\tau^A = X^{AA^\p}\xi_{A^\p}),
\ee
from which, under the choice of the gauge $\xi^T_{A^\p} = \{1,G\}$, one  
gets the constraints
\be{gg}
\tau^1 = wG+u,~~\tau^2 = vG+p .
\ee

Any generating twistor function (\ref{genf}) gives rise to a pair of the CEE 
solutions. To obtain the first one, one resolves the algebraic constraint
\be{kerr}
\Pi(G,wG+u,vG+p) = 0 
\ee
with respect to $G$ and comes then to a complex (and a {\it multivalued} 
in general) field $G(u,v,w,p)$ (or $G(x,y,z,t)$) which {\it for any $\Pi$ 
identically satisfies the CEE}. Secondly, one differentiates the 
function (\ref{genf}), resolves then again the arising constraint 
\be{sing}
P\equiv \frac{d\Pi}{dG} = 0
\ee
with respect to $G$ and substitutes the resulting field $G(u,v,w,p)$ into 
(\ref{genf}). In this way one comes to a function of coordinates 
$\Pi(G(X),\tau^1(X),\tau^2(X))\equiv\Pi(x,y,z,t)$ which again 
{\it identically satisfies the CEE}. 
It was shown in~\cite{Eik} that the two classes above presented 
{\it exhaust the solutions to CEE}.

Geometrically, any twistor field ${\bf W}(X) =\{G,\tau^1,\tau^2\}$ defines a 
{\it null geodesic congruence}, i.e. a congruence of rectilinear light-like  
rays on $\bf M$. For twistors obtained from the constraint (\ref{kerr}) 
this congruence is known to be always {\it shear-free} (SFC) and, 
moreover, any SFC can be constructed in this way (this is the content of the 
so called {\it Kerr theorem}~\cite{PR,DKS}). 

Conditions (\ref{kerr}),(\ref{sing}) considered together specify the 
{\it caustic locus} of the related congruence so that after elimination of 
$G$ the resulting equation
\be{eqmotion}
\Pi(u,v,w,p)\equiv \Pi(x,y,z,t) = 0 
\ee
(with the function $\Pi$ always obeying the CEE, see above) {\it determines 
the shape and the temporal evolution of caustics} which can be enormously 
complicated~\cite{Trishin,Sing,Pavlov}. On the other hand, the same 
condition defines the {\it singular locus} of the associated Maxwell and 
Yang-Mills fields~\cite{Kasan,Joseph}, as well as of the curvature field of 
correspondent Kerr-Schild metrics~\cite{DKS,Wilson}. Therefore, we can 
identify {\it particles as singularities~\footnote{At least for the case 
when the singular locus is bounded in 3D space}, common to the 
electromagnetic and the other fields defined by any solution of the GCRE (or, 
equivalently, by any SFC)}~\cite{Trishin,Kasan,Joseph}. 

Note that these singularities possess some of the properties of real quantum 
particles due, in particular, to the {\it overdetermined} structure of the GCRE. 
In fact, according to the {\it theorem of charge quantization} proved in
~\cite{Vestnik,Sing}, any bounded and isolated singularity of electromagnetic 
field associated with solutions of GCRE has the value of electric charge either 
zero or integer multiple to some minimal, {\it elementary} one, precisely 
to the charge of static axisymmetrical {\it Kerr solution} with a 
ring-like singularity~\cite{AD,GR,Kasan,Protvino}. According to observations of 
Carter~\cite{Carter} and Newman~\cite{Newman1}, the gyromagnetic ratio for  
the Kerr-like singularities is exactly the same as that for Dirac fermions. 
Numerous examples of SFC with complicated structure and dynamics of singular 
loci has been presented in our works~\cite{Kasan,Trishin,Protvino,Joseph,Pavlov}.    
 
From general viewpoint, we come in this way to a peculiar picture of physical World 
whose main elements are the congruence of the ``primodial light'' rays 
(invisible for an ordinary observer) -- the flow of {\it Prelight}~\cite{Number,
PIRT,Pavlov} -- and the matter ``born'' from this ``Ether-like'' yet Lorentz 
invariant flow at its focal points, i.e. at caustics. 
A consistent concept of physical Time as the 
{\it field evolutional parameter} along the rays of the Prelight flow has been 
also developed and discussed in our papers ~\cite{PIRT,Pavlov,Number,Levich} 
(see also section 3). 

Unfortunately, the above presented algebraic theory for interacting 
fields-particles is far from being realistic. Indeed, generically therein 
the particles-caustics are represented by a number of isolated 1D curves 
({\it strings}) nontrivially evolving in time but, as a rule, unstable 
in shape and size and radiating to infinity. Besides, one finds no correlation 
in temporal dynamics of different strings which could model their interaction. 
Finally, from the epistemological viewpoint, the real Minkowski subspace is by 
no means distinguished in the structure of $\mathbb{B}$-algebra: its 
elemens does't even form a subalgebra of $\mathbb{B}$.  

The last argument is of an especial importance in the framework of the 
algebrodynamics. In fact, no algebra is known whose {\it automorphism group} is 
isomorphic to the Lorentz group of Special Relativity. Therefore, in the 
algebrodynamical paradigm one is forced to admit space-time geometries  
different from the Minkowski one. By this, the full 4D {\it complex} vector 
space of the $\mathbb{B}$-algebra seems to be the most appropriate background 
for the algebric field-particle theory based  on the 
$\mathbb{B}$-differentiability conditions, i.e. on the GCRE. 

Complex extension $\mathbb{C}\bf M$ of the Minkowski space-time indeed arises  
repeatedly in General Relativity, in twistor and string theories. In 
particular, Newman et al.~\cite{Newman1,Newman2,Newmanetal} made use of the 
concept of complex space-time in order to obtain physically interesting solutions of 
electrovacuum Einstein-Maxwell system and to link together physical   
characteristics of corresponding particle-like singular sources (see the next 
section). In our recent paper~\cite{Levich} the algebrodynamics in the  
$\mathbb{C}\bf M$ space has been developed  on the base of the Newman's 
representation for ``virtual'' point charge ``moving'' along a complex world 
line in $\mathbb{C}\bf M$ and generating therein a congruence of complex 
``light-like rays''. The ``cut'' of this null congruence by the real Minkowski 
slice $\bf M$ gives there rise to a SFC with interesting physical properties, in 
particular with nonzero {\it twist}.  

Here again, however, one deals with a rather artificial and insufficient 
restriction of the basic structures onto $\bf M$. Alternatively, in the 
framework of the algebrodynamics in $\mathbb{C}\bf M$~\cite{Levich}, we have 
introduced the concept of the {\it observable} space-time -- the 6D subspace of 
the {\it complex null cone} (CNC) of a point-like ``observer'' $\bf O$ in 
$\mathbb{C}\bf M$. By this, any particle-like (caustic) element $\bf C$ which 
could be detected by $\bf O$ {\it necessarily lies on its CNC} and has the 
same value of the primary twistor field as in $\bf O$, so that the temporal 
dynamics of $\bf C$ and of $\bf O$ are strongly correlated (they ``interact''). 

On the other hand, the CNC has the topology $\mathbb{R}_+ \times {\bf S^3} 
\times {\bf S^2}$ (see section 3), so that it decomposes into a basic 4D space 
(in which the time interval should be identified with the 4D distance as in
~\cite{Montanus,Almeida}) and an orthogonal space of a 2-sphere which can be 
naturally treated as an internal {\it spin space}. 

Moreover, we'll see that an ensemble of identical point-like particles -- 
{\it duplicons} -- images of one and the same generating charge   
with correlated temporal evolution -- can be naturally constructed in this 
framework. We'll also discuss probable physical sense of the {\it quaternionic 
time}, {\it complex time} and of the {\it evolutionary curve} -- of notions 
inevitably arising in the considered algebrodynamical scheme and related,  
perhaps, to the origin of quantum uncertainty.

\section{Newman's charge in $\mathbb{C}{\bf M}$ and the set of its images -- duplicons} 
\label{charge}

Simplest variety of a shear-free null congruence (SFC) is formed by a bundle of light-like 
rays from a point particle moving along an arbitrary world line in the real 
$\bf M$ space~\cite{Kinnersley}. Electromagnetic field $F_{\mu\nu}$ 
can be then associated with such a SFC via integration of Maxwell equations 
with additional requirement on the field strength $F_{\mu\nu} k^\nu = 0$ to be 
orthogonal to the null 4-vector $k^\mu = \xi_A \xi_{A^\p}$ tangent to 
the congruence rays. Quite expectably, this field is just the Lienard-Wichert 
one. In the algebrodynamical approach, the field of exactly the same type can be obtained 
directly via second derivatives of the congruence field $G$~\cite{GR,Kasan,Protvino});   
however, this field necessarily corresponds to an {\it elementary (unit) value of generating 
charge}~\footnote{Owing to the overdetermined structure of the GCRE (or of SFC equations), 
see section 1}. Notice that in the static case the SFC generated by the charge at rest is 
radial and gives rise to the ``quantized'' Coulomb field and to the 
Reissner-N\"ordstrem metrics. 

Newman et al. proposed~\cite{Lind,Newman0} a generalization of this construction via 
consideration of a point-like charge ``moving'' in the complex extention 
$\mathbb{C}\bf M$ of $\bf M$ (in fact, even the case of the {\it curved} 
space-time was studied). Then on the real slice $\bf M$ of $\mathbb{C}\bf M$ 
one obtains a more complicated SFC with nonzero {\it twist} which, in the 
particular case of ``virtual'' charge at rest, gives rise to the metric and 
electromagnetic field of the well-known {\it Kerr-Newman} electrovacuum 
solution with a {\it ring-like singularity} (``Kerr's ring''). 

Consider now the Newman's construction in twistor terms~\cite{Burin2,Burin} and 
its generalization in the framework of algebrodynamics. Let a point singularity 
``moves'' along a complex world line $Z_\mu = {\wh Z}_\mu (\tau),~\mu=0,1,2,3$ 
where the parameter $\tau \in \mathbb{C}$ plays the role of ``complex time'' 
(further on we'll consider its own ``evolutionary law'' $\tau=\tau (s),~s\in 
\mathbb{R}$). For Cartezian and spinor coordinates on $\mathbb{C}\bf M$ 
we''ll use the following representation similar to (\ref{restr}):
\be{restrC}
Z = \{Z^A_{~B}\} = 
\left(
\begin{array}{cc}
u & w \\
p & v 
\end{array}\right)=
\left(
\begin{array}{cc}
~z_0 - iz_3 & -iz_1 - z_2 \\
-iz_1 + z_2 & ~z_0 + iz_3 
\end{array}\right)
\ee
in which all of the $z_\mu$ (as well as $u,v,w,p$) are now complex-valued. 
Incidence relation (\ref{inc}) takes then the form 
\be{incC}
\tau = Z\xi~~(\tau^A = Z^A_{~B} \xi^B,~~~\tau^1=wG+u,~\tau^2=vG+p), 
\ee
where the quantities $\{G,\tau^1,\tau^2\},~G\equiv \xi^2/\xi^1$ constitute the 
projective twistor in $\mathbb{C}\bf M$. 
If now a point  $Z\in \mathbb{C}\bf M$ is separated from the charge $\wh Z$ by 
the null interval,
\be{null}                      
S:=\det \Vert Z-{\wh Z}(\sigma) \Vert = (u-{\wh u}(\sigma))(v-{\wh v}(\sigma)) - 
(w-{\wh w}(\sigma))(p-{\wh p}(\sigma)) = 0, 
\ee
then one can find from (\ref{null}) the field $\sigma = \sigma(Z)=
\sigma(u,v,w,p)$ and, consequently, the position of the charge ${\wh Z}(Z)
\equiv {\wh Z}(\sigma(Z))$ which ``influences'' the point $Z$. This means that 
{\it twistor field} is the same both at $Z$ and at ${\wh Z}$. Indeed, in account 
of (\ref{null}) the linear system of equations
\be{linsyst}
(Z-{\wh Z}(\sigma))\xi = 0
\ee
in a unique way fixes the ratio of spinor components $G=\xi^2/\xi^1$. System 
(\ref{linsyst}) then reads as
\be{eqtau}
\tau \equiv Z\xi = {\wh Z}\xi, 
\ee
so that all of the twistor components are indeed equal at $Z$ and ${\wh Z}$.

Notice that the complex parameter $\sigma$, as a function of the point of 
observation $Z$, always satisfy the eikonal equation. Now in components 
system (\ref{linsyst}) takes the form
\be{fund}
\left\{
\begin{array}{lcl}
(u-{\wh u}(\sigma))+(w-{\wh w}(\sigma))G&=&0\\
(p-{\wh p}(\sigma))+~(v-{\wh v}(\sigma))G&=&0  
\end{array}
\right.
\ee
and demonstrates that the field $G$ is {\it indefinite} on the world line of the 
charge itself (and depends on the direction in its vicinity). On the other hand,
writing out (\ref{eqtau}) in the form 
\be{eqtauD}
\left\{
\begin{array}{rcl}
\tau^1&=&{\wh u}(\sigma) + {\wh w}(\sigma)G\\
\tau^2&=&{\wh p}(\sigma) + {\wh v}(\sigma)G
\end{array}
\right.
\ee
and eliminating from here the parameter $\sigma$, one concludes on the  
functional dependence of three twistor components, i.e.
\be{kerr2}
\Pi(G,\tau^1,\tau^2) = 0,
\ee
which, as in the real case (see (\ref{kerr})), demonstrates that fundamental null 
congruence of ``rays'' generated by the Newman's charge is indeed 
{\it shear-free} for any its world line (or, differently, for any 
respective twistor function $\Pi$).

Let us recall now the fundamental property of {\it multivaluedness} of the GCRE 
solutions or, equivalently,  -- of the solutions to the SFC equations (\ref{kerr}),  
or (\ref{fund}) in the considered case of the Newman's congruence. 
Precisely, for any fixed point $Z=\{u,v,w,p\}\in \mathbb{C}\bf M$ one obtains 
a countable (finite or infinite) set of solutions $\{G,\sigma\}$ of 
the system (\ref{fund}) so that, generically, {\it there exists 
a great number of continious branches -- ``modes''} -- of the twistor field functions 
${\bf W}(Z)$, of the eikonal field $\sigma(Z)$, as well as a great number of 
respective shear-free {\it subcongruences}, superposing at each point $Z$. 
Notice that general concept of multivalued fields has been discussed in~\cite{Pavlov}.
   
We see now that {\it any} complex ray of {\it any} mode of a SFC originates from 
a generating charge at some its position ${\wh Z}(\sigma)$ which, therefore, is 
a {\it focal point} of the congruence. Thus, the World line of the Newman's 
generating charge can be itself considered as the {\it focal line}. 

One should distinguish between focal points with indefinite value of the main spinor 
field $G$ and those of caustics (envelopes of the congruence rays) at which, generically, 
all of the field modes are well defined but two or more of them merge in one. 
For the Newman's SFC cauistic points, in account of (\ref{null}), satisfy the 
condition 
\be{caust}
S^\p = \frac{dS}{d\sigma} ={\wh u}^\p (v-{\wh v}(\sigma)) + {\wh v}^\p (u-{\wh u}(\sigma))
- {\wh w}^\p (p-{\wh p}(\sigma)) - {\wh p}^\p (w-{\wh w}(\sigma))= 0 . 
\ee
Locus of ``hupercaustic'' points -- {\it cusps} -- at which at least some 
{\it three} of the modes amalgamate -- are analogously defined by the condition  
\be{cusp}
S^{\p\p}:=\frac{d^2 S}{d\sigma^2} = {\wh u}^{\p\p}(v-{\wh v})+ 
{\wh v}^{\p\p}(u-{\wh u})- {\wh w}^{\p\p}(p-{\wh p}) - {\wh p}^{\p\p}
(w-{\wh w})-2({\wh u}^\p {\wh v}^\p  - {\wh w}^\p {\wh p}^\p) = 0  
\ee
and, finally, for the strongest singularities ({\it hypercusps}) one has the 
constraint $S^{\p\p\p}= ... = 0$. 
More stronger singularities have degenerate codimension space and generically  
exist in 4-dimensional $\mathbb{C}\bf M$ space only as isolated points.  

From (\ref{null}) and (\ref{caust}) we easily find that at any focal point at 
least {\it two} modes merge together so that for any $Z = {\wh Z}(\lambda)$, 
setting $\sigma = \lambda$, one gets identically $S = S^\p \equiv 0$. However, 
in the special case of a {\it null} World line defined as follows:  
\be{nullworld}
\det \Vert ({\wh Z}^\p \Vert \equiv {\wh u}^\p {\wh v}^\p  - {\wh w}^\p 
{\wh p}^\p \equiv 0, 
\ee
at any point $Z = {\wh Z}(\lambda)$ on the World line with $\sigma = \lambda$ 
one can easily prove that more stronger conditions for singularity are satisfied: 
$S=S^\p=S^{\p\p}=S^{\p\p\p}=0$, so that now {\it four} modes merge in one  
and are indefinite at the focal point itself. Further on we assume that 
{\it the World line of generating charge is null}. 

Note that the considered equation of the null cone (\ref{null}) in the case of 
the real space-time $\bf M$, in particular in the framework of electrodynamics, 
is known as the {\it retardation equation}. With real Cartesian coordinates 
$\{x,y,z,t\}$ and natural parametrization ${\wh t}(s) = s = \Re(\sigma)$ it 
takes the form
\be{retard}
c^2 (t-s)^2 - (x-x(s))^2 - (y-y(s))^2 - (z-z(s))^2 = 0,
\ee
where the charge moves along a real world line $\{x(s),y(s),z(s)\}$. If the 
velocity $v$ of the charge is assumed to be smaller than the light one, 
$v<c$, then equation (\ref{retard}) can be proved (see, e.g.,~\cite{Jackson}) 
to have always two roots only one from which has $s\le t$ and satisfies thus 
the causality requirement.  This fixes a unique position of the generating 
charge {\it in the past} which forms the field (electromagnetic Lienard-Wichert 
field in particular) at the observation point $\{x,y,z,t\}$ propagating to it  
rectilinearly with the speed of light. If the latter point belongs to the 
trajectory of the charge itself, equation (\ref{retard}) has only trivial 
solution $s=t$ (i.e., the retardation time is equal to zero, the observation 
point coincides with the ``influence'' one). 

Situation changes drastically in the complex space-time. In fact, now a whole 
set of solutions $\{\sigma_N\}$ of the complex null cone (CNC) equation 
(\ref{null}) does exist (for any point $Z$) which specify a (great) number of  
``influence points'', i.e. of the {\it images} of one and the same charge each 
forming its own mode of twistor field at $Z$ and perceiving, therefore, by the 
observer at $Z$ as a representative of the whole {\it ensemble of 
identical but distinguishable (via individual position and dynamics) point charges 
-- {\bf duplicons}}. 

Notice that total {\it number} of duplicons does not depend, as a rule, on the 
observation point. However, a realistic ``physical'' observer $\bf O$ should be 
{\it material} and, therefore, in the simplest idelized case can be identified 
with a focal point itself. Then for the observation point one has 
$Z = {\wh Z}(\lambda)$, with varying parameter $\lambda \in \mathbb{C}$ playing 
the role of the {\it complex proper time} of an ``elementary'' observer and 
manifesting iself as an evolutional parameter (see below).  Now four or 
two of the duplicons (the number depends on the fact is the World line null 
(\ref{nullworld}) or not) merge together forming the focal point of the 
observer $\bf O$ himself whereas the other $N-4$ (or $N-2$) duplicons are 
perceived by $\bf O$ as an ensemble of external identical point particles which 
are disposed and ``move'' with respect to $\bf O$ in the {\it observable 6D 
space-time} defined by the CNC equation (\ref{null}). CNC geometry and its 
reduction to the 4D physical space-time will be studied in the next section.

\section{Complex time and 6D physical geometry of the complex null cone}
\label{CNC}

In the algebrodynamics on the real $\bf M$ fundamental twistor field 
inevitably arises from the primary structure of $\mathbb{B}$-differentiable 
functions-fields. By this, physical time $t$ manifests itself as a {\it field 
evolutional parameter}. Indeed, the incidence relation (\ref{inc}) is invariant 
under a 1-parameter group of translations of coordinates $X=X^+\in \bf M$ 
along the rays of correspondent null congruence:
\be{transl}
x_a \to x_a + n_a t, ~(a=1,2,3);~~x_0 \to x_0 + t, ~~n_a = \xi^+ \sigma_a \xi / 
\xi^+ \xi, ~~\vec n^2 \equiv 1.
\ee
Presence of this symmetry demonstrates that the twistor field {\it propagates} 
from any initial point with universal velocity $v=c=1$ along the directions 
specified by vector $\vec n$ (and locally defined by the field itself) or, 
equivalently, is {\it preserved} in value along these in the 4D-sense. 
Just the parameter $t$ along the rays can be treated as local physical time 
(see correspondent discussion in~\cite{Pavlov,Number,Levich}).

When one generalizes the scheme to the complex background, the same 
considerations lead to a more complicated and intriguing notion of time. Indeed, 
in $\mathbb{C}\bf M$ corresponding symmetry group of translations is 
$2\mathbb{C}$-parametric (the coordinates can vary across the so called 
{\it complex $\alpha$-plane}~\cite{PR}). We'll, however, start from an 
alternate representation of this field symmetry what for decompose the space 
$\mathbb{C}\bf M$ of biquaternion algebra into two {\it real quaternionic} 
(4D Euclidean) spaces represented by $2\times 2$ {\it unitary} matrices 
$U$ and $V$:
\be{decomp}
Z = U + i V, ~~~U^+ =U^{-1},~V^+ = V^{-1}.                                   
\ee
The first of these subspaces, say $\bf U$, may be considered as the main 
coordinate space whereas the second one $\bf V$ will then play the role of 
the {\it evolutional parameter space}. Precisely, from the incidence condition 
(\ref{incC}) and the decomposition (\ref{decomp}) one gets, 
after some equivalent transformations, the following {\it translation law}:
\be{evolQ}
U = U^{(0)} + V*N,
\ee
where the quaternion $U^{(0)}$ of initial coordinates is algebraically 
expressed via the twistor field components, quaternion 
$V=v_0+i v_a\sigma_a$ defines four real evolutional parameters 
$\{v_0,v_a\}$ (``quaternionic time''), and the {\it unit} quaternion 
$N=1+i n_a\sigma_a$ (with the vector $\vec n$ of the same form as in the real 
case (\ref{transl})) specifies local directions for the {\it translations 
preserving twistor field}.  

Interestingly, the whole law (\ref{evolQ}) does not contain any residual of 
initial complex structure since all of the matrices therein are unitary and 
$(*)$ can be considered thus as multiplication in the {\it real quaternion 
algebra} $\mathbb{H}$. In the components, fundamental translation law 
(\ref{evolQ}) for four coordinates $\{x_\mu\},~\mu =0,1,2,3$ of the main 
{\it Euclidean} space $U\in {\bf E^4},~U=x_0+ix_a\sigma_a$ has the following 
form:
\be{translQ}
\Delta x_a =  n_a \Delta v_0 + \varepsilon_{abc} n_b \Delta v_c, ~~~
\Delta x_0 =  n_a \Delta v_a ,
\ee
so that any of translations satisfies one complex (two real) constraints 
$\Sigma (\Delta z_\mu)^2 = 0$, or in the components: 
\be{cncQ} 
\left\{
\begin{array}{lcl}
(\Delta x_0)^2 + (\Delta x_1)^2 + (\Delta x_2)^2 + (\Delta x_3)^2 = 
(\Delta v_0)^2 + (\Delta v_1)^2 + (\Delta v_2)^2 + (\Delta v_3)^2 \equiv (\Delta t)^2, \\
(\Delta x_0)(\Delta v_0) +(\Delta x_1)(\Delta v_1) +(\Delta x_2)(\Delta v_2) +
(\Delta x_3)(\Delta v_3) = 0, 
\end{array}
\right.
\ee
fixing the structure of the {\it complex null cone}.  

Notice that in the particular case when three of the parameters 
$\Delta v_a$ vanish and the fourth one acquires the meaning of physical time 
$\Delta v_0 = t$,  three ``space'' coordinates $\{x_a\}$ are translated just as 
in the former Minkowski case (\ref{transl}), i.e. rectilinearly and with 
fundamental ``light'' velocity whereas for additional Euclidean coordinate one 
has $\Delta x_0=0$, so that it is ``freezed'' and looses its dynamical sense.

Generically, the translation law (\ref{translQ}) is much more complicated 
and rich in structure. The presence of four evolutional parameters makes the 
field dynamics essentially uncertain since the {\it order} of continious 
change of the parameters is not determined by the structure. However, we 
can naturally assume that the role of local physical time interval is now 
performed by the total Euclidean distance $(\Delta t)^2 =\Sigma (\Delta 
x_\mu)^2$ passing by the field in the coordinate space. In account of (\ref
{cncQ}) this is equal to corresponding distance in the parameter space and twice 
smaller than that in the whole complex space with natural metric $\Sigma \vert 
z_\mu \vert^2$. 

Thus, in the generic complex case local physical time can be related to the 
translations preserving the value of basic twistor field and to the whole 
distance to which the field propagates via these translations. In the latter 
aspect such treatment of physical time is similar to that proposed in the works 
of H.Montanus~\cite{Montanus} and J.Almeida~\cite{Almeida}. They considered 
the model of space-time as a 4D Euclidean space $\bf E^4$ with fourth coordinate  
being identified with local {\it proper time} of a particle of matter, and with 
local {\it coordinate time interval} being the metric interval covered by the 
latter (in natural units). By this, by analogy with the old ideas of F.Klein, 
Yu.Rumer et al., it is assumed that {\it all the material formations move in 
$\bf E^4$ with constant in modulus fundamental velocity} 
$\Sigma \left({\Delta x_\mu}/{\Delta t}\right)^2 = 1~(=c^2).$ 
From here it follows immegiately that {\it 3-dimensional velocity of all 
material particles may be only less than or equal to the fundamental one}. 

We see thus that in the context of successive biquaternionic 
electrodynamics, under the reduction of the primary ``observable'' 
geometry of the complex null cone of ``elementary'' observer, there 
arises a {\it model of the 4D Euclidean space-time} similar to that 
postulated in the approach of Montanus-Almeida. However, restriction of 
the 3-velocities' values by the fundamental value holds in the algebrodynamics  
only for propagation of ``prematerial'' twistor field. As for the 
particle-like formations represented here by the focal and by the 
(hyper-) caustic points, this property will be identically fulfiled only  
for their {\it absolute} velocities, and only if the World line is null.   
For {\it relative} (``observable'') velocities situation is more 
complicated (see below). On the other hand, the structure of   
cone includes, apart of the main Euclidean ``physical'' space $\bf U$, a  
2D orthogonal space $\bf V$ of the 2-sphere which could be treated (at least 
at the focal points of the particle's location) as the space of  
directions of its {\it spin vector}. 

Unfortunately, the scheme based on the concept of the 4D Euclidean space-time 
suffers from evident inconsistencies with Special Relativity. Indeed, though 
formally  the first constraint in (\ref{cncQ}) can be represented in the form of  
the Minkowski-like interval $ds^2:= dx_0^2 = dt^2 - dx_1^2 - dx_2^2 - dx_3^2$,  
there are no grounds for the proper time interval $ds$ to remain invariant 
under the 4D rotations, as well as for the coordinate time $dt$ to transform 
according to the Lorentz group. In the biquaternionic algebrodynamics 
situation is much more satisfactory since, under the transformations of 
the automorphism group $SO(3,\mathbb{C}$ of the $\mathbb{B}$-algebra, the 
proper time represented by the zeroth coordinate, indeed remains invariant.   
Even by this, however, transformations of the spacial coordinates and 
coordinate time differ from the canonical Lorentz ones. This problem 
demands further investigation. 

Let us recall now that in the algebrodynamical approach we have also to 
solve the above-mentioned problem of indefinite character of field evolution 
related to the presence of three additional free evolutional parameters $v_a$ 
or, on the other hand, -- the problem of {\it ordering} of the 4D 
``quaternionic time''. This last difficulty can be partially overcomed in a 
rather successive way. Note that under arbitrary field-preserving translations 
with four parameters $\{v_\mu\}$ the particle-like formations (focal points, 
caustics etc.) are not preserved and, generally, dissappear from the main 
space $\bf E^4$. 

In particular, if one takes all $v_\mu = 0$ to get the 
field-particle distribution at an {\it initial} instant of time, say at $t=0$, 
then, generically, one will find the space $\bf E^4$ {\it empty}, i.e. free 
from any focal points. Indeed, the latters are {\it holomorphically} defined as 
the points of a complex World line $Z = {\wh Z} (\sigma)$, and all four 
equations necessary for coordinates of these points to be {\it unitary} can't be 
satisfied for any complex value of the parameter $\sigma$. 

Thus, the problem is to concord the structure of the main quaternionic Euclidean space 
$\bf U$ and the ``real quaternionic'' type of the field evolution law 
(\ref{translQ}) with the basic holomorphic structure of the null congruence and 
its particle-like singularities. To do this, let us return back to the 
above introduced {\it observable} space-time $\bf O$, i.e. to the 6D space-time 
of the complex null cone of an elementary observer $\bf O$ which is identified 
with a focal point and moves along its World line in synchro with the process 
of field evolution. On this {\it relative} subspace modelling physical 
space-time, we are now going to consider the evolutionary law for coordinates 
$\bf U$ preserving the structure of basic equations (\ref{null}) and 
(\ref{caust}) together, i.e. {\it preserving both the value 
of twistor field and the condition of caustic}. This extended ``field-caustic 
preserving'' automorphism has the following $1\mathbb{C}$-parameter holomorphic 
form: 
\be{evolcaust}
Z \to  Z + \tau (Z-{\wh Z}(\sigma)), ~~\tau \in \mathbb{C}.    
\ee
This means that caustics reproduce themselves along the complex straight lines 
ending (starting) at a focal point, and the parameter $\tau$ along these 
should be considered as the only physical evolutional parameter -- the 
``complex time''. In the real case such situation could correspond to 
a ``signal'' propagating rectilinearly with fundamental velocity. Complex 
structure, however, is not ordered, so there are infinitely many ``ways'' 
connecting some two points on such a line. Moreover, one is even not aware 
which of them is the emitter, and which -- the receiver point. Note that 
the concept of two- (or multi- in general) dimensional time and the resulting  
problem of ordering of physical events have been considered in 
a number of works, in particular by Sakharov~\cite{Sakharov}. In the works of  
Yefremov~\cite{Yefrem1,Yefrem2} just the structure of the automorphism                                                       
group $SO(3,\mathbb{C})$ of the biquaternion algebra has been exploited, and 
at its base the concept of {\it three-dimensional time} was put forward. Under 
special orthogonality requirements this was assumed to reduce to the ordinary  
physical one-dimensional time. 

{\it Uncertainty} of the field evolution process related to the complex 
nature of time is weaker than that for the primary ``quaternionic time'' but 
still too severe for usual physical interpretation, at least if one  
adheres to the paradigm of classical field theory. 

To concord the concept of complex time with classical representations, one 
could assume, in addition to the holomorphic structure of the GCRE (or of the 
SFC equations), the existence of some ``evolutionary curve'' 
$\tau = \tau(s),~s \in \mathbb{R}$ on the complex plane which establishes the 
order of passing of different ``states'' in the main ``physical'' space $\bf U$. 
Such a curve as well as the World line of generating charge can be enormously  
complicated, possess the points of self-intersection etc. Because the 
observer has no information about the form of this curve, the future for him 
seems to be indefinite. However, a sort of {\it probabilities} can be 
introduced for different ``continuations'' of the evolutionary curve, and this 
procedure can open the way for explaination of quantum phenomena and for   
revealing the links of algebrodynamics and the Feynman's version of quantum 
mechanics in particular. These problems will be discussed elsewhere. 

There is evidently a local correspondence between the ``evolutionary time'' $\tau$, 
the parameter $\sigma$ of the World line  and, say, the zeroth coordinate of 
the observer which can be treated as the ``complex proper time''. In account of 
the quaternionic representation of the complex null cone  of field evolution 
(\ref{cncQ}) we can interpret then the real part $\Delta x_0$ of 
``complex time increment'' $\Delta z_0$ as an evolutional parameter in the 
main coordinate space $\bf U$ (``translational'' proper time), whereas its 
{\it imaginary part} $\Delta v_0$ -- as that in the orthogonal spin space 
(``rotational'' proper time). Existence of the evolutionary curve implies 
also direct local connection between the increment of its parameter $s$ and 
corresponding interval of the coordinate time $\Delta t$. Explicit formulas 
can be easily presented. 

Let us consider now in more details the above mentioned process of 
emission/reception of a ``signal-caustic'' by ``charges -- focal points''. 
Let the ``interacting'' charges -- duplicons ${\wh Z}(\lambda),~{\wh Z}(\sigma),
~\sigma\ne \lambda$ are linked by the caustic. This corresponds 
to a solution of the joint system of equations (\ref{null}) and (\ref{caust}), 
with substitution $Z = {\wh Z}(\lambda)$ instead of taking an arbitrary 
observation point $Z$. This system has a {\it discrete} 
set of pairs of solutions $\{\lambda,\sigma\}$ any of which correspond to 
two fixed {\it instants of complex time} of emission/reception of the caustic 
signal and, respectively, to two positions of the charges involved into the 
process. 

By this, if the moment of observation $\lambda$ 
``lies in the future'' with respect to the ``moment of influence'' $\sigma$, 
i.e. corresponds to a greater value of the monotonically increasing parameter 
$s$ of the evolutionary curve, than the observer ${\wh Z}(\lambda)$ would seem 
to {\it receive} the signal from the charge at ${\wh Z}(\sigma)$, and the 
latter would {\it emit} the signal towards ${\wh Z}(\lambda)$. 
The converse situation is evident. However, one should not forget 
that, in essence, this is {\it one and the same} charge {\it interacting with 
itself in its own past or future}. 

Thus, all the {\it events} in complex space-time are well-defined and predetermined 
by the only structure of the World line of the generating charge (the focal line 
of the fundamental congruence). These are: the total number of charges -- 
duplicons (``observed'' by anyone of them; relative coordinates of duplicons
{\it permanently} correlated with those of the observer through equal values 
of the primary twistor field); and, finally, a set of points on the 
World line at which this correlation between a pair of duplicons  
is stronly {\it amplified} (the charges are involved into the emission/reception 
process). However, the precise picture of evolution still remains indefinite 
and should be completed by additional information about the true form of the 
evolutionary curve (extremely complicated and exceptional from the mathematical 
viewpoint). 
                  
A microscopic object (a focal point) is detectable for an idealized ``elementary'' 
observer only at some discrete moments of time at which the caustic line links 
them together. Near these, {\it relative} instantaneous velocity of the object 
is easily proved to be equal to its {\it absolute} one and, for the considered 
case of a {\it null} World line, is always less or equal to the fundamental 
velocity (the speed of light).
These conclusions seem to hold also for a ``classical'' 
macroscopic object which is nearly permanently linked with a similar 
``classical'' observer by a {\it net} of caustic lines. A detailed description  
(in the two utmost cases -- quantum (microscopic) and classical (macroscopic)) 
of the correlated pairs ``object - observer'' exchanging by caustic ``signals'' 
will be presented elsewhere.

\end{document}